\documentclass[traditabstract]{aa} % for the abstract without structuration
\usepackage{graphicx}
\usepackage{txfonts}
\usepackage{natbib}
\usepackage{longtable}
\usepackage{amsmath}
\bibpunct[]{(}{)}{;}{a}{}{,}

\begin{document}

   \title{Investigation of the rotational spectrum of CD$_3$OD and an astronomical search
          toward IRAS 16293$-$2422\thanks{Transition 
          frequencies from this and earlier work are given as supplementary material. 
          We  also provide quantum numbers, uncertainties, and residuals between 
          measured frequencies and those calculated from the final set of 
          spectroscopic parameters. The data are available at Centre de Données astronomiques de
          Strasbourg (CDS) via anonymous 
          ftp to cdsarc.u-strasbg.fr (130.79.128.5) or via 
          http://cdsweb.u-strasbg.fr/cgi-bin/qcat?J/A+A/}}

   \author{
           V.~V. Ilyushin\inst{1}
           \and
           H.~S.~P. M{\"u}ller\inst{2}
           \and
           J.~K. J{\o}rgensen\inst{3}
           \and
           S. Bauerecker\inst{4}
           \and
           C. Maul\inst{4}
           \and
           R. Porohovoi\inst{1}
           \and
           E.~A. Alekseev\inst{1}
           \and
           O. Dorovskaya\inst{1}
           \and
           F. Lewen\inst{2}
           \and
           S. Schlemmer\inst{2}
           \and
           R.~M. Lees\inst{5}
           }

   \institute{Institute of Radio Astronomy of NASU, Mystetstv 4, 61002 Kharkiv, Ukraine\\
              \email{ilyushin@rian.kharkov.ua}
              \and
              I.~Physikalisches Institut, Universit{\"a}t zu K{\"o}ln,
              Z{\"u}lpicher Str. 77, 50937 K{\"o}ln, Germany\\
              \email{hspm@ph1.uni-koeln.de}
              \and
              Niels Bohr Institute, University of Copenhagen, {\O}ster Voldgade 5$-$7, 1350 Copenhagen K, Denmark 
              \and
              Institut f{\"u}r Physikalische und Theoretische Chemie, 
              Technische Universit{\"a}t Braunschweig, Gau{\ss}str. 17, 38106 Braunschweig, Germany
              \and
              Department of Physics, University of New Brunswick, Saint John, NB E2L 4L5, Canada
              }

   \date{Received 06 Jun 2023 / Accepted 26 Jun 2023}
 
%%%%%%%%%%%%%%%%%%%%%%%%%%%%%%%%%%%%%%%%%%%%%%%%%%%%%%%%%%%%%%%%%%%%%%%%%%%%%%%%%%%%%%%%%
%%%%%%%%%%%%%%%%%%%%%%%%%%%%%%%%%%%%%%%%%%%%%%%%%%%%%%%%%%%%%%%%%%%%%%%%%%%%%%%%%%%%%%%%%
\abstract
%%%%%%%%%%%%%%%%%%%%%%%%%%%%%%%%%%%%%%%%%%%%%%%%%%%%%%%%%%%%%%%%%%%%%%%%%%%%%%%%%%%%%%%%%
%%%%%%%%%%%%%%%%%%%%%%%%%%%%%%%%%%%%%%%%%%%%%%%%%%%%%%%%%%%%%%%%%%%%%%%%%%%%%%%%%%%%%%%%%

\abstract{
Solar-type prestellar cores and protostars frequently display  large amounts of deuterated 
organic molecules and, in particular, high relative abundances of doubly and triply deuterated 
isotopologs. Recent findings on CHD$_2$OH and CD$_3$OH toward IRAS 16293$-$2422 suggest that 
even fully deuterated methanol, CD$_3$OD, may be detectable as well. However, searches for 
CD$_3$OD are hampered in particular by the lack of intensity information from a spectroscopic 
model. The objective of the present investigation is to develop a spectroscopic model of 
CD$_3$OD in low-lying torsional states that is sufficiently accurate to facilitate searches 
for this isotopolog in space. 
We carried out a new measurement campaign for CD$_3$OD involving two spectroscopic laboratories 
that covers the 34~GHz$-$1.1~THz range. A torsion-rotation Hamiltonian model based on the 
rho-axis method was employed for our analysis. Our resulting model describes the ground and 
first excited torsional states of CD$_3$OD well up to quantum numbers $J \leqslant 51$ and 
$K_a \leqslant 23$. We derived a line list for radio-astronomical observations from this model 
that is accurate up to at least 1.1~THz and should be sufficient for all types of 
radio-astronomical searches for this methanol isotopolog. This line list was used to search 
for CD$_3$OD in data from the Protostellar Interferometric Line Survey of IRAS 16293$-$2422 
obtained with the Atacama Large Millimeter/submillimeter Array. While we found several emission 
features that can be attributed largely to CD$_3$OD, their number is still not sufficiently 
high enough to establish a clear detection. 
Nevertheless, the estimate of 2$\times 10^{15}$~cm$^{-2}$ derived for the CD$_3$OD column density 
may be viewed as an upper limit that can be compared to column densities of CD$_3$OH, CH$_3$OD, and CH$_3$OH. 
The comparison indicates that the CD$_3$OD column density toward IRAS 16293$-$2422 is in line 
with the enhanced D/H ratios observed for multiply deuterated complex organic molecules.
}


\authorrunning{V.~V. Ilyushin et al.}
\titlerunning{Rotational spectroscopy of CD$_3$OD}

\maketitle
\hyphenation{For-schungs-ge-mein-schaft}

%\onecolumn

%%%%%%%%%%%%%%%%%%%%%%%%%%%%%%%%%%%%%%%%%%%%%%%%%%%%%%%%%%%%%%%%%%%%%%%%%%%%%%%%%%%%%%%%%
%%%%%%%%%%%%%%%%%%%%%%%%%%%%%%%%%%%%%%%%%%%%%%%%%%%%%%%%%%%%%%%%%%%%%%%%%%%%%%%%%%%%%%%%%
\section{Introduction}
\label{intro}
%%%%%%%%%%%%%%%%%%%%%%%%%%%%%%%%%%%%%%%%%%%%%%%%%%%%%%%%%%%%%%%%%%%%%%%%%%%%%%%%%%%%%%%%%
%%%%%%%%%%%%%%%%%%%%%%%%%%%%%%%%%%%%%%%%%%%%%%%%%%%%%%%%%%%%%%%%%%%%%%%%%%%%%%%%%%%%%%%%%

Methanol, CH$_3$OH, is among the most abundant polyatomic molecules in the interstellar medium (ISM) 
as evidenced by its early radio astronomical detection \citep{det_CH3OH_1970}. It is observed 
both in its solid state and gas phase toward star-forming regions \citep[e.g.,][]{Herbst2009} 
and is an important product of the chemistry occurring on the icy surfaces of dust grains 
\citep[e.g.,][]{Tielens1982,Garrod2006}. As a slightly asymmetric rotor, whose excitation 
is strongly dependent on kinetic temperature, methanol presents a useful diagnostic tool 
for evaluating the physical conditions prevailing in star-forming regions \citep{Leurini2004}.
Due to its ubiquity in the ISM, methanol is often taken as a reference for studies 
of the chemistry of more complex organic molecules \citep[e.g.,][]{Jorgensen2020}.

The ubiquity and high abundance of interstellar methanol make this molecule suitable for studying 
the degree of deuteration, which is considered as an indicator of the evolution of low-mass star-forming 
regions \citep{deuteration_2005,deuteration_2007,deuteration_2018}. Not only has singly deuterated methanol been detected in the ISM, but 
also doubly and triply deuterated  as well. The singly deuterated methanol isotopologs, 
CH$_3$OD \citep{det_CH3OD_1988} and CH$_2$DOH \citep{det_CH2DOH_1993}, were detected first. 
Some time later, \citet{det_CHD2OH_2002} observed CHD$_2$OH toward IRAS 16293$-$2422, 
followed by \citet{det_CD3OH_2004} detecting CD$_3$OH toward the same object. 
In addition, CHD$_2$OH was also found toward several other low-mass protostars 
\citep{more_CHD2OH_2006,deuterated_CH3OH_2019} and, most recently, in a prestellar core 
\citep{CHD2OH_prestellar_2023}.

Multiply deuterated isotopic species frequently appear to be overabundant in comparison to the D/H ratio 
inferred from the singly and non-deuterated species (see e.g., results for doubly deuterated isotopologs 
of methyl cyanide (CHD$_2$CN; \citealt{CH3CN_PILS_2018}), methyl formate (CHD$_2$OCHO; \citealt{D2-MeFo_2019}), 
and dimethyl ether (CHD$_2$OCH$_3$; \citealt{D2-DME_2021}) toward the low-mass protostellar system IRAS 16293$-$2422), 
which may reflect their formation processes at low temperatures \citep{deuteration_in_ice_2014}.
Recently revisited abundances of CHD$_2$OH and CD$_3$OH toward IRAS 16293-2422 employing the Protostellar Interferometric Line Survey (PILS) data \citep{CHD2OH_catalog_2022,CD3OH_rot_2022} also demonstrate this overabundance, suggesting that a search for fully deuterated isotopolog of methanol, CD$_3$OD, would be promising and timely.

The rotational spectrum of CD$_3$OD had already been observed in the lab in the 1950s  
in the context of other methanol isotopologs, in particular, to determine the molecular structure 
\citep{MeOH-isos_1-0_I_1955}. \citet{CD3OD_rot_1972} published an account of the rotational spectrum of CD$_3$OD 
in the microwave region. Additional rotational transition frequencies in the millimeter and/or submillimeter region 
and with infrared \citep{CD3OD_rot-IM_2004} or microwave accuracies were reported later 
\citep{3isos_rot_1992,CD3OD_rot_1998,CD3OD_rot-K_2004,CD3OD_rot-K_2006}. 
Torsional transition frequencies were provided in two studies very recently 
\citep{CD3OD_rot_etc-IM_2021,CD3OD_rot_etc-IM_2022}; these publications also contain  some millimeter and 
submillimeter assignments. The rovibrational spectrum of CD$_3$OD beyond the torsional manifold was also 
investigated in some studies, with \citet{CD3OD_MIR_2022} being the most recent one dealing with the COD bending 
fundamental at 775~cm$^{-1}$.

The goal of our present investigation is to develop a spectroscopic model of the CD$_3$OD isotopolog in 
low-lying torsional states which is sufficiently accurate to provide reliable calculations 
of line positions and linestrengths for astronomical searches for CD$_3$OD in the ISM. 
New measurements were carried out to extend the covered frequency range up to 1.1~THz. 
The obtained new data were combined with previously published far infrared measurements 
to form the final dataset involving the rotational quantum numbers up to $J$ = 51 and $K$ = 23. 
A fit within experimental error was obtained for the ground and first excited torsional states of 
the CD$_3$OD molecule using the so-called rho-axis-method. 

We generated a line list that was based on our present results, which we applied in a search for CD$_3$OD 
in the Atacama Large Millimeter/submillimeter Array (ALMA) data of the Protostellar Interferometric Line Survey 
of the deeply embedded protostellar system IRAS 16293$-$2422 \citep{PILS_2016}. 
While we did not detect CD$_3$OD confidently, a number of emission lines that can be attributed to CD$_3$OD 
(or at least to a large part of it) suggest that a detection is within the reach of ALMA, for instance, 
by targeting IRAS~16293-2422 through deep observations at lower frequencies -- where line confusion may be less problematic.

The rest of the manuscript is organized as follows. Section~\ref{exptl} provides details on 
our laboratory measurements. The theoretical model, spectroscopic analysis, 
and fitting results are presented in Sections~\ref{spec_backgr} and \ref{lab-results}. 
Section~\ref{astrosearch} describes our astronomical observations and the results of our search 
for CD$_3$OD, while Section~\ref{conclusion} gives the conclusions of our investigation.

%%%%%%%%%%%%%%%%%%%%%%%%%%%%%%%%%%%%%%%%%%%%%%%%%%%%%%%%%%%%%%%%%%%%%%%%%%%%%%%%%%%%%%%%%
%%%%%%%%%%%%%%%%%%%%%%%%%%%%%%%%%%%%%%%%%%%%%%%%%%%%%%%%%%%%%%%%%%%%%%%%%%%%%%%%%%%%%%%%%
\section{Experimental details}
\label{exptl}

%%%%%%%%%%%%%%%%%%%%%%%%%%%%%%%%%%%%%%%%%%%%%%%%%%%%%%%%%%%%%%%%%%%%%%%%%%%%%%%%%%%%%
\subsection{Rotational spectra at IRA NASU}
%%%%%%%%%%%%%%%%%%%%%%%%%%%%%%%%%%%%%%%%%%%%%%%%%%%%%%%%%%%%%%%%%%%%%%%%%%%%%%%%%%%%%
%%%%%%%%%%%%%%%%%%%%%%%%%%%%%%%%%%%%%%%%%%%%%%%%%%%%%%%%%%%%%%%%%%%%%%%%%%%%%%%%%%%%%

The measurements of the CD$_3$OD spectrum at the Institute of Radio Astronomy (IRA) of NASU 
were performed in the frequency ranges of 34.5$-$184~GHz and 234$-$420~GHz using an automated synthesizer 
based millimeter wave spectrometer \citep{Alekseev2012}. This instrument belongs to a class of 
absorption spectrometers and uses a set of backward wave oscillators (BWO) to cover 
the frequency range from 34.5 to 184~GHz, allowing for further extension to the 234$-$420 GHz range 
with the help of a solid state tripler from Virginia Diodes, Inc. (VDI). 
The frequency of the BWO probing signal is stabilized by a two-step frequency multiplication of 
a reference synthesizer in two phase-lock-loop stages. A commercial sample of CD$_3$OD was used 
and all measurements were carried out at room temperature with sample pressures providing linewidths 
close to the Doppler limited resolution (about 2~Pa). Due to the high rate of D/H exchange at the OH group 
in CD$_3$OD, the recorded spectrum contains numerous lines belonging to the CD$_3$OH isotopolog. 
These lines do not pose any problem since they may be easily distinguished using the results of our 
recent study of the CD$_3$OH spectrum \citep{CD3OH_rot_2022}. 
Estimated uncertainties for measured line frequencies were 10~kHz, 30~kHz, 50~kHz, and 100~kHz depending 
on the observed signal-to-noise ratios (S/N).

%%%%%%%%%%%%%%%%%%%%%%%%%%%%%%%%%%%%%%%%%%%%%%%%%%%%%%%%%%%%%%%%%%%%%%%%%%%%%%%%%%%%%
%%%%%%%%%%%%%%%%%%%%%%%%%%%%%%%%%%%%%%%%%%%%%%%%%%%%%%%%%%%%%%%%%%%%%%%%%%%%%%%%%%%%%
\subsection{Rotational spectra at the Universit{\"at} zu K{\"o}ln}
%%%%%%%%%%%%%%%%%%%%%%%%%%%%%%%%%%%%%%%%%%%%%%%%%%%%%%%%%%%%%%%%%%%%%%%%%%%%%%%%%%%%%
%%%%%%%%%%%%%%%%%%%%%%%%%%%%%%%%%%%%%%%%%%%%%%%%%%%%%%%%%%%%%%%%%%%%%%%%%%%%%%%%%%%%%

%%% done

The measurements at the Universit{\"a}t zu K{\"o}ln were carried out at room temperature 
using a 5~m long single path Pyrex glass cell of 100~mm inner diameter and equipped with 
high-density polyethylene windows. The cell was filled with 1.5~Pa CD$_3$OD and refilled 
after several hours because of the pressure rise due to minute leaks. 
We utilized three VDI frequency multipliers driven by 
Rohde \& Schwarz SMF~100A synthesizers as sources and a closed cycle liquid He-cooled 
InSb bolometer (QMC Instruments Ltd) as detector to cover frequencies between 370 and 
1095~GHz almost entirely; a small gap near 750~GHz occurred because of a strong water line. 
Other water lines or low power, especially at the edges, limited the sensitivity 
in some frequency regions. Frequency modulation was used throughout. 
The demodulation at $2f$ causes an isolated line to appear close to a second derivative 
of a Gaussian. Additional information on this spectrometer system is available in 
\citet{CH3SH_rot_2012}. We were able to achieve uncertainties of 5$-$10~kHz for very 
symmetric lines with good S/N, as demonstrated in recent studies on 
excited vibrational lines of CH$_3$CN \citep{MeCN_up2v4eq1_etc_2021} or on isotopic 
oxirane \citep{c-C2H4O_rot_2022,c-C2H3DO_rot_2023}. Uncertainties of 10~kHz, 30~kHz, 
50~kHz, 100~kHz, and 200~kHz were assigned, depending on the observed S/N and on 
the frequency range.

%%%%%%%%%%%%%%%%%%%%%%%%%%%%%%%%%%%%%%%%%%%%%%%%%%%%%%%%%%%%%%%%%%%%%%%%%%%%%%%%%%%%%%%%%
%%%%%%%%%%%%%%%%%%%%%%%%%%%%%%%%%%%%%%%%%%%%%%%%%%%%%%%%%%%%%%%%%%%%%%%%%%%%%%%%%%%%%%%%%
\section{Spectroscopic properties of CD$_3$OD and our theoretical approach}
\label{spec_backgr}
%%%%%%%%%%%%%%%%%%%%%%%%%%%%%%%%%%%%%%%%%%%%%%%%%%%%%%%%%%%%%%%%%%%%%%%%%%%%%%%%%%%%%%%%%
%%%%%%%%%%%%%%%%%%%%%%%%%%%%%%%%%%%%%%%%%%%%%%%%%%%%%%%%%%%%%%%%%%%%%%%%%%%%%%%%%%%%%%%%%

%% I left this part as it may serve as a good starting point 
%% with maybe modest adjustments; some rewording may be useful nevertheless

Fully deuterated methanol, CD$_3$OD, is a nearly prolate top ($\kappa \approx -0.959$) with 
a rather high coupling between internal and overall rotations in the molecule ($\rho \approx 0.82$) 
and the torsional potential barrier $V_3$ of about 362~cm$^{-1}$. The torsional problem corresponds 
to an intermediate barrier case \citep{RevModPhys.31.841} with the reduced barrier $s = 4V_3/9F$ $\sim$10.9, 
where $F$ is the rotation constant of the internal rotor. In comparison with the parent isotopolog, 
CD$_3$OD has significantly smaller rotational parameters: $A \approx 2.17$~cm$^{-1}$, $B \approx 0.630$~cm$^{-1}$, 
$C \approx 0.598$~cm$^{-1}$ in CD$_3$OD versus $A \approx 4.25$~cm$^{-1}$, $B \approx 0.823$~cm$^{-1}$, 
$C \approx 0.792$~cm$^{-1}$ in CH$_3$OH \citep{XU:2008305}. Thus, we expect that rotational levels with 
higher $J$ and $K$ values will be accessible in a room temperature experiment for CD$_3$OD compared to CH$_3$OH.

As the theoretical approach in the present study, we employ the so-called rho-axis-method (RAM), 
which has proven to be the most effective approach so far in treating torsional large amplitude motions 
in methanol-like molecules. The method is based on the work of \citet{Kirtman:1962}, 
\citet{CH_D3OH_D_rot_1968}, and \citet{Herbst:1984} and takes its name from the choice 
of its axis system \citep{Hougen:1994}. In the rho-axis-method, the $z$ axis is coincident with the $\rho$ vector, 
which expresses the coupling between the angular momentum of the internal rotation $p_{\alpha}$ and 
that of the global rotation $J$. We employed the RAM36 code \citep{Ilyushin:2010,Ilyushin:2013} 
that was successfully used in the past for a number of near prolate tops with rather high $\rho$ and $J$ values 
(see e.g., \citet{Smirnov:2014}, \citet{Motiyenko:2020}, \citet{Zakharenko:2019}) and, in particular, for 
the CD$_3$OH isotopolog of methanol \citep{CD3OH_rot_2022}. The RAM36 code uses the two-step diagonalization procedure of \citet{Herbst:1984} and in the current study, we keep 31 torsional basis functions at the first diagonalization step and 11 torsional basis functions at the second diagonalization step.

The labeling scheme after the second diagonalization step is based on an eigenfunction composition but, in contrast 
to our CD$_3$OH study \citep{CD3OH_rot_2022}, is not limited to searching for a dominant eigenvector component only. 
Since methanol is a nearly symmetric prolate top ($\kappa \approx -0.98$), in which the angle between the RAM a-axis 
and the principal-axis-method (PAM) a-axis is only 0.07$^{\circ}$, it is assumed that the RAM a-axis in methanol 
is suitable for $K$ quantization and that eigenvectors can be unambiguously assigned using dominant components. 
And indeed, searching for a dominant eigenvector component worked well in the case of the CD$_3$OH isotopolog 
\citep{CD3OH_rot_2022}, for which the angle between the RAM $a$-axis and the PAM $a$-axis is only 0.14$^{\circ}$ 
and the asymmetry parameter ($\kappa \approx -0.977$) is nearly the same  as in CH$_3$OH ($\kappa \approx -0.982$). 
In CD$_3$OD, however, the angle between the RAM $a$-axis and PAM $a$-axis is 0.68$^{\circ}$ and $\kappa \approx -0.959$, 
and it appeared that starting from $J \approx 25$, some eigenvectors do not have a dominant eigenvector component 
that would allow for unambiguous labeling. Therefore, we have employed a combined labeling scheme. 
First, we search for a dominant eigenvector component ($\geq 0.8$) and if it exists, the level is labeled according 
to this dominant component. If such component is absent then we search for similarities in basis-set composition 
in torsion-rotation eigenvectors belonging to the previous $J$ value and assign the level according to the highest degree of similarity found. 
The general idea assumes that for a given pair of $K_a$, $v_t$ values, the torsion–rotation eigenfunctions vary slowly 
when $J$ changes by one, and this slow change should appear as a high degree of similarity in the eigenvector compositions 
of the states corresponding to the same $K_a$, $v_t$ , and adjacent $J$ values. This approach allows one to transfer 
a given $K_a$-label from lower $J$ values, where it can be determined easily (either from eigenvector composition using dominant component or from energy-ordering considerations), to higher $J$ values, which are characterized by extensive basis-set mixing. 
The details of this approach to $K$-labeling for torsion–rotation energy levels in low-barrier molecules can be found 
in \citet{ILYUSHIN2004}.

The energy levels are labeled in our fits and predictions by the free rotor quantum number, $m$, 
the overall rotational angular momentum quantum number, $J$, and a signed value of $K_a$, which is the axial 
$a$-component of the overall rotational angular momentum, $J$. In the case of the A symmetry species, 
the $+/-$ sign corresponds to the so-called "parity" designation, which is related to the A1/A2 symmetry species 
in the group $G_6$ \citep{Hougen:1994}. The signed value of $K_a$ for the E symmetry species reflects the fact 
that the Coriolis-type interaction between the internal rotation and the global rotation causes the $|K_a| > 0$ levels 
to split into a $K_a > 0$ level and a $K_a < 0$ level. We also provide $K_c$ values for convenience, but they are 
simply recalculated from the $J$ and $K_{a}$ values, $K_{c} = J - |K_{a}|$ for $K_{a} \geq 0$ 
and $K_{c} = J - |K_{a}| + 1$ for $K_{a} < 0$. The $m$ values 0, $-$3, 3 / 1, $-$2, 4 
correspond to A/E transitions of the $\varv_{\rm t} = 0$, 1, and 2 torsional states, respectively.

%%%%%%%%%%%%%%%%%%%%%%%%%%%%%%%%%%%%%%%%%%%%%%%%%%%%%%%%%%%%%%%%%%%%%%%%%%%%%%%%%%%%%%%%%
%%%%%%%%%%%%%%%%%%%%%%%%%%%%%%%%%%%%%%%%%%%%%%%%%%%%%%%%%%%%%%%%%%%%%%%%%%%%%%%%%%%%%%%%%
\section{Spectroscopic results}
\label{lab-results}
%%%%%%%%%%%%%%%%%%%%%%%%%%%%%%%%%%%%%%%%%%%%%%%%%%%%%%%%%%%%%%%%%%%%%%%%%%%%%%%%%%%%%%%%%
%%%%%%%%%%%%%%%%%%%%%%%%%%%%%%%%%%%%%%%%%%%%%%%%%%%%%%%%%%%%%%%%%%%%%%%%%%%%%%%%%%%%%%%%%

We started our analysis from the results of \citet{CD3OD_rot-K_2006}, where the dataset, consisting of 
488 $\varv_{\rm t} = 0$ and 182 $\varv_{\rm t} = 1$ microwave transitions, ranging up to $J_{\rm max} = 25$ 
and $K_{\rm max} = 14$, was fit with 56 parameters of the RAM Hamiltonian, and a weighted standard 
deviation of 3.13 was achieved. Since the BELGI code \citep{Kleiner:2010} was used in the previous study 
\citep{CD3OD_rot-K_2006}, we refit the available dataset with the RAM36 program 
\citep{Ilyushin:2010,Ilyushin:2013} as the first step. The resulting fit was the starting point of our present analysis.

New data were assigned starting with the Kharkiv measurements. First, the search for the $\varv_{\rm t} = 2$ 
rotational transitions was performed with success, and further assignments were made in parallel 
for the three torsional states of CD$_3$OD  $\varv_{\rm t} = 0$, 1, and 2. Submillimeter wave and 
THz measurements from Cologne were assigned subsequently, based on our new results. 
Whenever it was possible, we replaced the old measurements from \citet{CD3OD_rot-K_2006} and 
references therein with the more accurate new ones. 
At the same time, we decided to keep in the fits two measured values for the same transition from the 
Kharkiv and Cologne spectral recordings in that part of the frequency range where the measurements 
from the two laboratories overlap (370$-$420~GHz). A rather good agreement within the experimental uncertainties 
was observed for this limited set of duplicate new measurements. Finally, at an advanced stage of our analysis, 
the FIR data from \citet{CD3OD_rot_etc-IM_2021} were added to the fit.

The assignment process was performed in a usual bootstrap manner, with numerous cycles of refinement 
of the parameter set while the new data were gradually added. In parallel, a search of the optimal set of 
RAM torsion–rotation parameters was carried out, and it finally became evident  that the $\varv_{\rm t}=2$ 
torsional state poses some problems with fitting. The strong influence of intervibrational interactions 
arising from low lying small amplitude vibrations in CD$_3$OD, which then propagate down through numerous 
intertorsional interactions, is a possible explanation for these problems. We encountered similar problems 
with CD$_3$OH \citep{CD3OH_rot_2022}. 
There we decided to limit our fitting attempts mainly to the ground and first excited torsional states. 
Taking into account that one of our goals was to provide reliable predictions for astrophysical searches 
of interstellar CD$_3$OD, we adopted an analogous decision in the current case of the CD$_3$OD investigation. 
Thus, at the final stage of model refinement, we limited our fitting attempts mainly to the ground and 
first excited torsional states of CD$_3$OD. 
Only the lowest three $K$ series for the A and E species in $\varv_{\rm t}=2$ were retained in the fits in order 
to get a better constraint of the torsional parameters in the Hamiltonian model. These $\varv_{\rm t}=2$ $K$ levels 
should be affected least by the intervibrational interactions arising from low lying small amplitude vibrations. 
In the case of CD$_3$OD, this corresponds to $K = -1, 2, 3$ for the E species in $\varv_{\rm t}=2$ and 
to $K = -4, 0, 4$ for the A species.

It should be noted that at the final stage of preparation of this manuscript, the new study of 
\citet{CD3OD_rot_etc-IM_2022} appeared. This study emphasized the transitions involving the 
second excited torsional state levels. Taking into account that our efforts are essentially 
concentrated on the ground and first excited torsional states of CD$_3$OD, we decided not to include any data from 
\citet{CD3OD_rot_etc-IM_2022} in the present analysis even those involving the lowest three $K$ series 
of levels for the A and E species in $\varv_{\rm t}=2,$ which were retained in the fits otherwise. 
Submillimeter wave transitions appeared from \citet{CD3OD_rot_etc-IM_2022} were also not included in our present fits 
since they are within the range of our current measurements, with our measurements being more precise. 
We intend to include the data from \citet{CD3OD_rot_etc-IM_2022} at the next stage of our investigation of CD$_3$OD, 
when we will try to fit transition higher in $\varv_{\rm t}$ and try to model the intervibrational interactions 
arising from low lying small amplitude vibrations in CD$_3$OD. With this aim in mind, new measurements of the CD$_3$OD 
IR spectrum between 500 and 1200~cm$^{-1}$ were carried out at the Technische Universit{\"at} Braunschweig, 
which we plan to consider in our future analyses of the CD$_3$OD spectrum.

Our final CD$_3$OD dataset for the purposes of this paper involves 4337 FIR and 10001 microwave line frequencies that correspond, due to blending, to 16259 transitions with $J_{\rm max} = 51$. Due to the duplication in measurements mentioned above, the number of unique transitions incorporated in the fit is  somewhat lower at 15135. A Hamiltonian model consisting of 117 parameters provided a fit with a weighted root mean square (RMS) deviation of 0.71 
which was selected as our "best fit" for this paper. The 117 molecular parameters from our final fit are given in 
Table~\ref{tbl:ParametersTable} (Appendix A). The numbers of the terms in the model distributed between the orders 
$n_{\rm op} = 2$, 4, 6, 8, 10, 12 are 7, 22, 45, 34, 8, and 1 respectively, which is consistent with the limits of 
determinable parameters of 7, 22, 50, 95, 161, and 252 for these orders, as calculated from the differences between 
the total number of symmetry-allowed Hamiltonian terms of order $n_{\rm op}$ and the number of symmetry-allowed 
contact transformation terms of order $n_{\rm op} - 1$, when applying the ordering scheme of \citet{Nakagawa:1987}. 
The final set of the parameters converged perfectly in all three senses: (i) the relative change in the weighted RMS deviation 
of the fit at the last iteration was about $\sim$10$^{-7}$ ; (ii) the corrections to the parameter values generated 
at the last iteration are less than $\sim$10$^{-3}$ of the calculated parameter confidence intervals;  and
(iii) the changes generated at the last iteration in the calculated frequencies are less than 1~kHz even for the FIR data.

\begin{figure}
\centering
\includegraphics[width=9cm,angle=0]{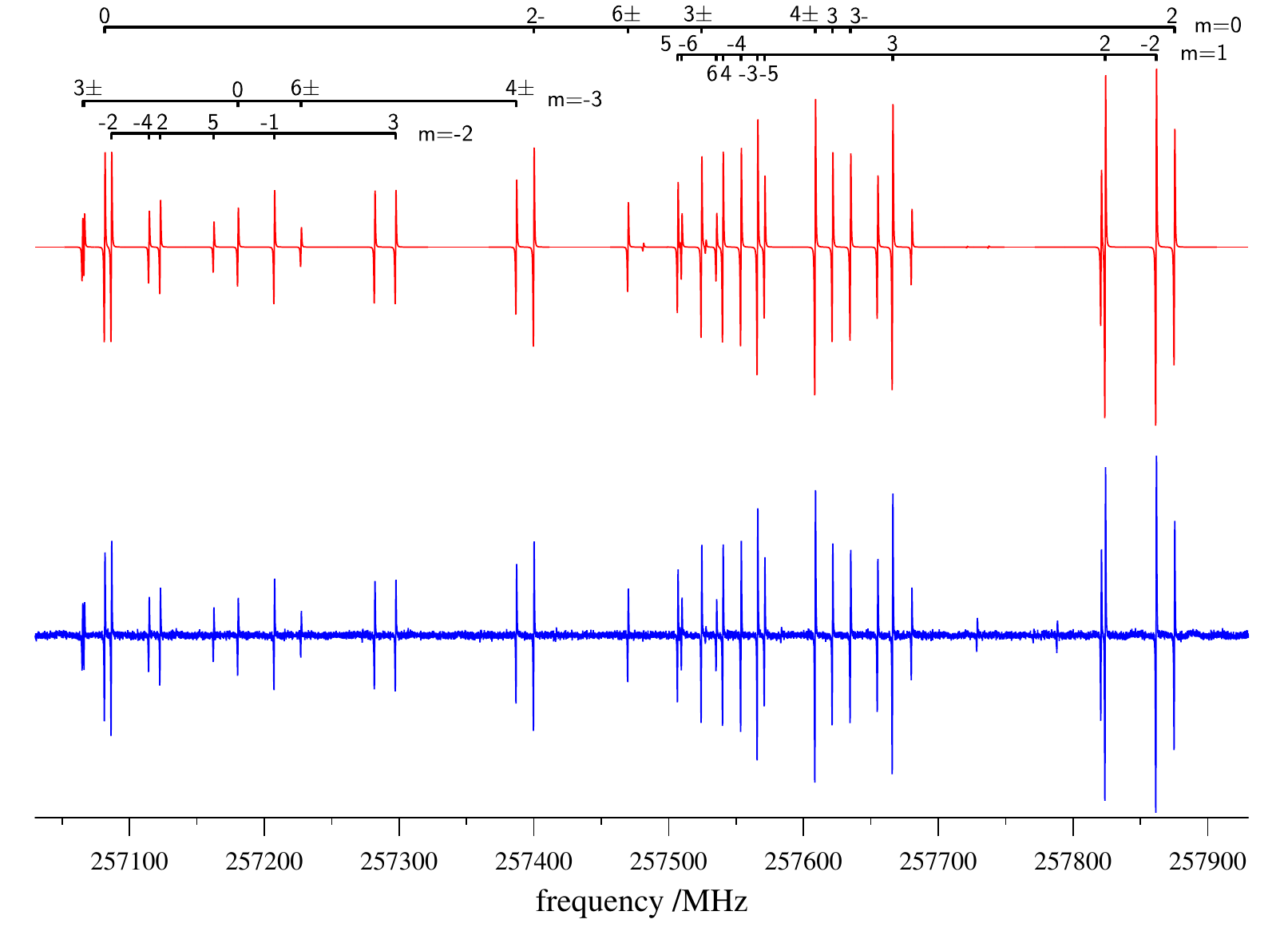}

\caption{Portion of the CD$_3$OD microwave spectrum dominated by $J = 7 \leftarrow 6$ $R$-branch of transitions 
in the 257.03$-$257.93~GHz range. The observed spectrum is shown in the lower panel and the calculated one in the upper one. 
The $K$ quantum numbers for the $m=0,1$ ($\varv_{\rm t} = 0$) and $m=-2, -3$ ($\varv_{\rm t} = 1$) $J = 7 \leftarrow 6$ 
transitions are given at the top.
It is seen that the experimental frequencies and the  intensity pattern are rather well reproduced by our model 
for the spectral features dominating this frequency range.}
\label{fig_MMW_Kharkiv}
\end{figure}

\begin{figure}
\centering
\includegraphics[width=9cm,angle=0]{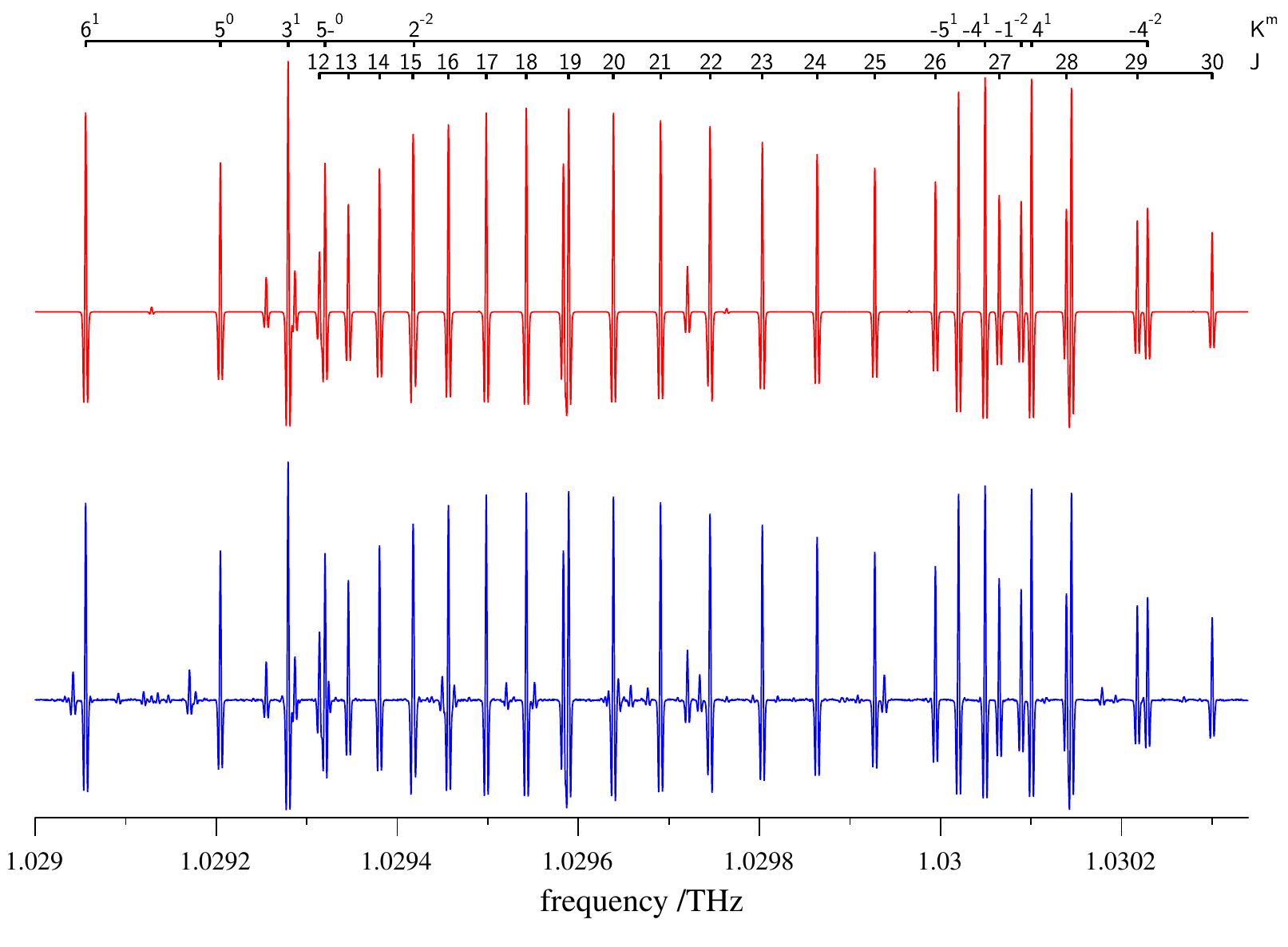}

\caption{Portion of the CD$_3$OD THz spectrum dominated by $m = 1$ $K= -12 \leftarrow -11$ $Q$-branch 
and $J = 28 \leftarrow 27$ $R$-branch transitions in the 1.029$-$1.030~THz range. The observed spectrum 
is shown in the lower panel and the calculated one in the upper one. The $J$ quantum numbers for the 
$m = 1$ $K= -12 \leftarrow -11$ $Q$-branch and the $K$ and $m$ quantum numbers ($K^m$) for the 
$J = 28 \leftarrow 27$ $R$-branch are given at the top.
It is seen that the experimental frequencies and the  intensity pattern are rather well reproduced 
by our model for the spectral features dominating this frequency range.}
\label{fig_THz_Cologne}
\end{figure}

A summary of the quality of this fit is given in Table~\ref{tbl:statisticInf}. The overall weighted RMS deviation of 0.71 
and the additional fact that all data groups are fit within experimental uncertainties (see the left part of 
Table~\ref{tbl:statisticInf} where the data are grouped by measurement uncertainty) seems satisfactory to us. 
If we consider the weighted RMS deviations for the data grouped by torsional state we will see a rather good agreement 
between our model and the experiment for all three torsional states as well as for the torsional fundamental band. 
Further illustration of our current understanding of the microwave spectrum of CD$_3$OD may be found at 
Figs.~\ref{fig_MMW_Kharkiv} and ~\ref{fig_THz_Cologne}. It is seen that our current model reproduces the observed microwave spectrum 
quite well both with respect to line positions and line intensities.  

%%%%%%%%%%%%%%%%%%%%%%%%%%%%%%%%%%%%%%%%%%%%%%%%%%%%%%%%%%%%%%%%%%%%%%%%%%%%%%%%%%%%%%%%%%

%%%%%%%%%%%%%%%%%%%%%%%%%%%%%%%%%%%%%%%%%%%%%%%%%%%%%%%%%%%%%%%%%%%%%%%%%%%%%%%%%%%%%%%%%

%%%%%%%%%%%%%%%%%%%%%%%%%%%%%%%%%%%%%%%%%%%%%%%%%%%%%%%%%%%%%%%%%%%%%%%%%%%%%%%%%%%%%%%%%%

%%%%%%%%%%%%%%%%%%%%%%%%%%%%%%%%%%%%%%%%%%%%%%%%%%%%%%%%%%%%%%%%%%%%%%%%%%%%%%%%%%%%%%%%%

Using the parameters of our final fit we calculated a list of CD$_3$OD transitions in the ground and 
first excited torsional states for astronomical observations. The dipole moment function of \citet{MEKHTIEV1999171} 
was employed in our calculations where the values for the permanent dipole moment components of CH$_3$OH were replaced 
by appropriate ones for CD$_3$OD $\mu_a = 0.867$~D and $\mu_b = 1.430$~D taken from \citet{MUKHOPADHYAY19981307}. 
The permanent dipole moment components were rotated from the principal axis system to the rho axis system 
of our Hamiltonian model. As in the case of CD$_3$OH \citep{CD3OH_rot_2022}, the list of CD$_3$OD transitions 
includes information on transition quantum numbers, transition frequencies, calculated uncertainties, 
lower state energies, and transition strengths.    
To avoid unreliable extrapolations far beyond the quantum number coverage of the available experimental dataset, 
we limited our predictions by $\varv_{\rm t} \leq 1$, $J \leq 55$ and $|K_{a}| \leq 25$. As already mentioned earlier, 
we label torsion-rotation levels by the free rotor quantum number, $m$, the overall rotational angular momentum 
quantum number, $J$, a signed value of $K_a$, and $K_c$. 
The calculations were done up to 1.33~THz. Additionally, we limit our calculations to transitions for which calculated uncertainties 
are less than 0.1~MHz. The lower state energies are given referenced to the $K_{a} = 0$ A-type $\varv_{\rm t} = 0$ level. 
We provide additionally the torsion-rotation part of the partition function $Q_{\rm rt}$(T) of CD$_3$OD calculated 
from first principles, that is, via direct summation over the torsion-rotational states. The maximum $J$ value is 
90 and $n_{\varv_{\rm t}} = 11$ torsional states were taken into account. The calculations, as well as 
the experimental line list from the present work, can be found in the online Supplementary material with this article 
and will also be available in the Cologne Database for Molecular Spectroscopy, \citep[CDMS,][]{Endres2016}.

%%%%%%%%%%%%%%%%%%%%%%%%%%%%%%%%%%%%%%%%%%%%%%%%%%%%%%%%%%%%%%%%%%%%%%%%%%%%%%%%%%%%%%%%%
%%%%%%%%%%%%%%%%%%%%%%%%%%%%%%%%%%%%%%%%%%%%%%%%%%%%%%%%%%%%%%%%%%%%%%%%%%%%%%%%%%%%%%%%%
\section{Astronomical search for CD$_3$OD}
\label{astrosearch}
%%%%%%%%%%%%%%%%%%%%%%%%%%%%%%%%%%%%%%%%%%%%%%%%%%%%%%%%%%%%%%%%%%%%%%%%%%%%%%%%%%%%%%%%%
%%%%%%%%%%%%%%%%%%%%%%%%%%%%%%%%%%%%%%%%%%%%%%%%%%%%%%%%%%%%%%%%%%%%%%%%%%%%%%%%%%%%%%%%%

The new spectroscopic calculations were used to search for CD$_3$OD in data from the 
Protostellar Interferometric Line Survey (PILS; \citealt{PILS_2016}). PILS represents an unbiased spectral survey 
of the Class~0 protostellar system IRAS~16293$-$2422 using the Atacama Large Millimeter/submillimeter Array 
covering the frequency range from 329 to 363~GHz. The data cover the region of IRAS~16293-2422 including 
its two primary components ``A'' and ``B'' that show abundant lines of complex organic molecules at an angular resolution 
of  $\sim$0.5$''$ and a spectral resolution of $\sim$0.2~km~s$^{-1}$. Toward a position slightly offset from 
the ``B'' component of the system, the lines are intrinsically narrow, making it an ideal hunting ground for new species 
and several complex organic molecules and their isotopologs have been identified there, 
including other deuterated isotopologs of CH$_3$OH such as CH$_2$DOH and CH$_3$OD \citep{deuteration_16293_2018}, 
CHD$_2$OH \citep{CHD2OH_catalog_2022}, and CD$_3$OH \citep{CD3OH_rot_2022}.

\begin{figure}
\resizebox{\hsize}{!}{\includegraphics{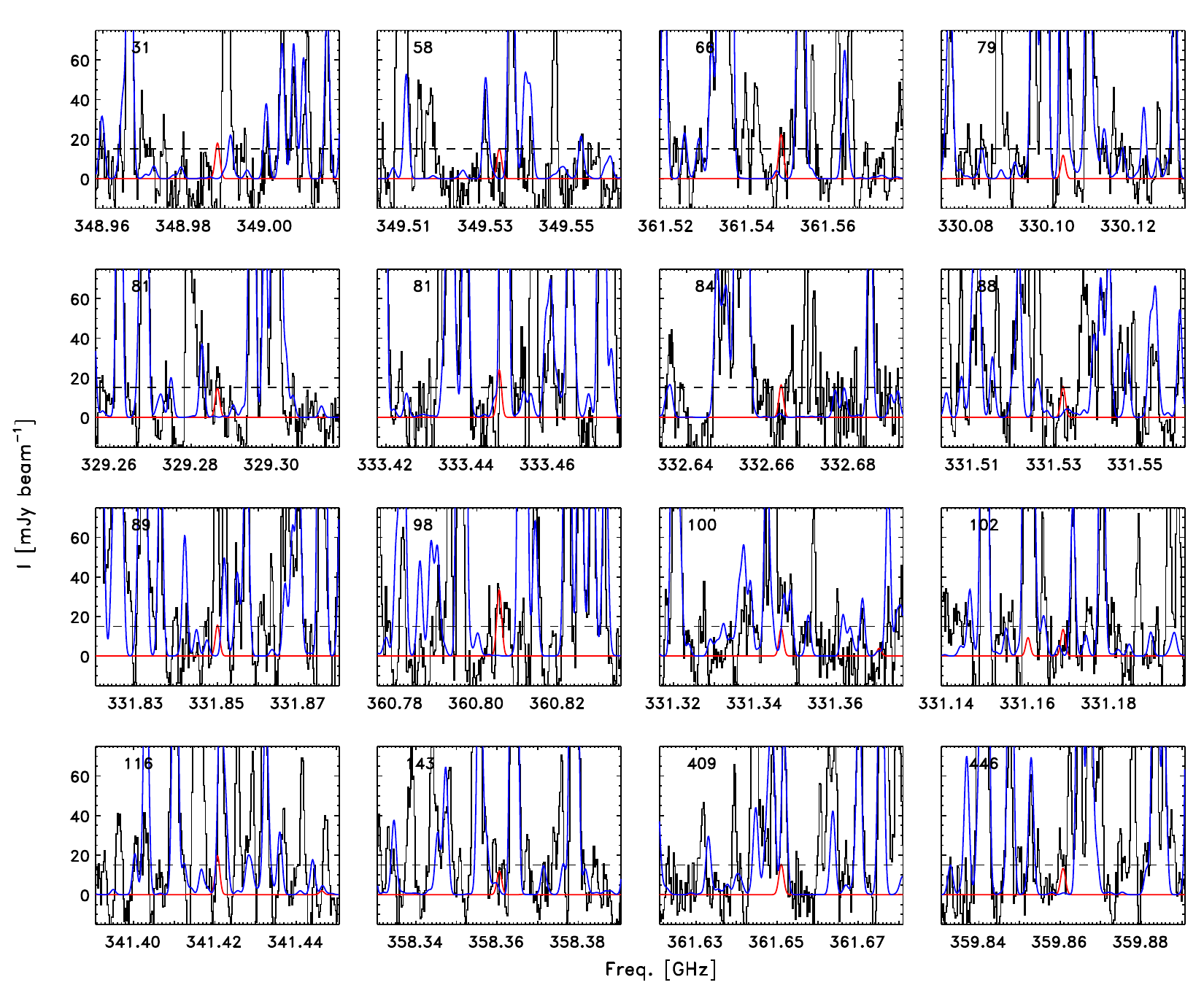}}
\caption{Sixteen lines predicted to be the strongest toward the position offset by 1 beam (0.5$''$) from 
IRAS16293B: the observations are shown in the black, the synthetic spectrum for CD$_3$OD in red, and other species 
identified as part of PILS in blue. The dashed line indicates the 3$\sigma$ RMS level of the data per 1~km~s$^{-1}$ 
(1 km~s$^{-1}$ is the typical FWHM line width of the lines toward this position).}
\label{obs}
\end{figure}

The search for CD$_3$OD was conducted following the approach applied in other papers from PILS: synthetic spectra 
are calculated assuming that the excitation of the molecule is characterized by local thermodynamical equilibrium, 
which is reasonable at the densities on the scales probed by PILS \citep{PILS_2016}. The source velocity offset relative 
to the local standard of rest is assumed to be 2.6~km~s$^{-1}$ and the line full width at half maximum (FWHM) 
is taken as 1~km~s$^{-1}$. With these assumptions the kinetic temperature and column density of the species are left 
as the two free parameters. For the search we assume a temperature of 225~K similar to that of CD$_3$OH \citep{CD3OH_rot_2022}. 
Figure~\ref{obs} shows the 16 lines predicted to be strongest for this temperature: the column density is taken to be 
the maximum possible without overproducing lines compared to the RMS noise level of the data with a value of 
2$\times 10^{15}$~cm$^{-2}$. As we can see, three to four lines match the observed spectral features at or slightly 
above the 3$\sigma$ level in the data. One line at 348.988~GHz is predicted at the 3$\sigma$ level but does not show 
any observed emission. However, that one is overlapping with the absorption part of a nearby, stronger, transition 
so that may not be significant. The quoted column density would correspond to an CD$_3$OD abundance of 0.02\% relative 
to the main isotopolog CH$_3$OH (the column density of that constrained by observations of optically thin transitions 
of CH$_3^{18}$OH). If all D/H substitutions would be considered equally probable, this in turn would imply a D/H ratio 
of about 12\% -- similar to what is measured for other multiple deuterated species and enhanced relative to the ratios 
measured from the singly deuterated variants (CH$_2$DOH and CH$_3$OD). While this makes the assignments of the three 
to four transitions plausible it is not possible to claim a solid detection based on so few lines. 
However, the analysis demonstrates that the detection of CD$_3$OD is within the reach of ALMA, for instance, 
by targeting IRAS~16293-2422 through deep observations at lower frequencies where line confusion may be less problematic.

%%%%%%%%%%%%%%%%%%%%%%%%%%%%%%%%%%%%%%%%%%%%%%%%%%%%%%%%%%%%%%%%%%%%%%%%%%%%%%%%%%%%%%%%%
%%%%%%%%%%%%%%%%%%%%%%%%%%%%%%%%%%%%%%%%%%%%%%%%%%%%%%%%%%%%%%%%%%%%%%%%%%%%%%%%%%%%%%%%%
\section{Conclusion}
\label{conclusion}
%%%%%%%%%%%%%%%%%%%%%%%%%%%%%%%%%%%%%%%%%%%%%%%%%%%%%%%%%%%%%%%%%%%%%%%%%%%%%%%%%%%%%%%%%
%%%%%%%%%%%%%%%%%%%%%%%%%%%%%%%%%%%%%%%%%%%%%%%%%%%%%%%%%%%%%%%%%%%%%%%%%%%%%%%%%%%%%%%%%

%% this maybe needs only slight adjustment; some rewording may be useful also

In this work, we performed a new study of the torsion-rotation spectrum of the CD$_3$OD isotopolog using 
a torsion-rotation RAM Hamiltonian. The new microwave measurements carried out in the broad 
frequency range from 34.5~GHz to 1.1~THz and transitions with $J$ up to 51 and $K_a$ up to 23 involving 
the $\varv_{\rm t}$ = 0, 1, 2 torsional states were assigned and analyzed. After revealing perturbations 
in the second excited torsional state of CD$_3$OD, presumably caused by the intervibrational interactions 
arising from low-lying small-amplitude vibrations in this molecule, we concentrated our 
efforts on refining the theoretical model for the ground and first excited torsional states 
only. A fit within the experimental uncertainties (weighted RMS deviation 0.71) was achieved 
for the dataset consisting of 4337 FIR and 10001 microwave line frequencies.

%% astronomical part need extensive rewriting; the CD3OH was erased

Based on our results, calculations of the ground and first excited torsional states were carried out and used 
in a search for CD$_3$OD spectral features in data from the ALMA PILS survey of the deeply 
embedded protostar IRAS 16293$-$2422. While three to four CD$_3$OD transitions match observed spectral features 
at or slightly above the 3$\sigma$ level in the data, it is not possible to claim a solid detection based on so few lines. 
Nevertheless, our analysis demonstrates that the detection of CD$_3$OD in IRAS~16293-2422 using ALMA is quite probable 
through deep observations at lower frequencies where line confusion may be less problematic. 
The upper column density limit of 2$\times 10^{15}$~cm$^{-2}$ for CD$_3$OD was derived based on the assumption of 
an excitation temperature of 225~K (taken similar to that of CD$_3$OH \citep{CD3OH_rot_2022}). 
Comparison with the CH$_3$OH main isotopolog (for which the column density is deduced from optically thin lines 
of CH$_3^{18}$OH) yields a CD$_3$OD/CH$_3$OH ratio as high as $\sim$0.02\%, thus implying that the fully deuterated methanol 
is in line with the enhanced D/H ratios observed for multiply deuterated complex organic molecules.

%%%%%%%%%%%%%%%%%%%%%%%%%%%%%%%%%%%%%%%%%%%%%%%%%%%%%%%%%%%%%%%%%%%%%%%%%%%%%%%%%%%%%%%%%
%%%%%%%%%%%%%%%%%%%%%%%%%%%%%%%%%%%%%%%%%%%%%%%%%%%%%%%%%%%%%%%%%%%%%%%%%%%%%%%%%%%%%%%%%

\begin{acknowledgements}
We acknowledge support by the Deutsche Forschungsgemeinschaft via the collaborative 
research center SFB~956 (project ID 184018867) project B3 as well as
the Ger{\"a}tezentrum SCHL~341/15-1 (``Cologne Center for Terahertz Spectroscopy''). 
The research in Kharkiv and Braunschweig was carried out under support of the Volkswagen foundation. 
The assistance of the Science and Technology Center in the Ukraine is acknowledged (STCU partner project P756). 
J.K.J. is supported by the Independent Research Fund Denmark (grant number 0135-00123B). 
R.M.L. received support from the Natural Sciences and Engineering Research Council of Canada. 
Our research benefited from NASA's Astrophysics Data System (ADS). 
This paper makes use of the following ALMA data: ADS/JAO.ALMA \# 2013.1.00278.S. 
ALMA is a partnership of ESO (representing its member states), NSF (USA) and NINS (Japan), together with 
NRC (Canada), MOST and ASIAA (Taiwan), and KASI (Republic of Korea), in cooperation with the Republic of Chile. 
The Joint ALMA Observatory is operated by ESO, AUI/NRAO and NAOJ.
\end{acknowledgements}

%%%%%%%%%%%%%%%%%%%%%%%%%%%%%%%%%%%%%%%%%%%%%%%%%%%%%%%%%%%%%%%%%%%%%%%%%%%%%%%%%%%%%%%%%
%%%%%%%%%%%%%%%%%%%%%%%%%%%%%%%%%%%%%%%%%%%%%%%%%%%%%%%%%%%%%%%%%%%%%%%%%%%%%%%%%%%%%%%%%

\bibliographystyle{aa} 
\bibliography{CD3OD}

%%%%%%%%%%%%%%%%%%%%%%%%%%%%%%%%%%%%%%%%%%%%%%%%%%%%%%%%%%%%%%%%%%%%%
%%%%%%%%%%%%%%%%%%%%%%%%%%%%%%%%%%%%%%%%%%%%%%%%%%%%%%%%%%%%%%%%%%%%%
%%%%%%    Table 1    %%%%%%%%%%%%%%%%%%%%%%%%%%%%%%%%%%%%%%%%%%%%%%%%
%%%%%%%%%%%%%%%%%%%%%%%%%%%%%%%%%%%%%%%%%%%%%%%%%%%%%%%%%%%%%%%%%%%%%
%%%%%%%%%%%%%%%%%%%%%%%%%%%%%%%%%%%%%%%%%%%%%%%%%%%%%%%%%%%%%%%%%%%%%
\newpage
\begin{table*}[]
\caption{\label{tbl:statisticInf} Overview of the dataset and the fit quality }
\begin{tabular}{lrl|lrl}
\hline
\multicolumn{3}{c|}{By measurement uncertainty}& \multicolumn{3}{c}{By torsional state} \\ 
\cline{1-6}
 
\multicolumn{1}{c}{Unc.$^a$} & \multicolumn{1}{c}{$\#^b$} & \multicolumn{1}{c|}{RMS$^c$} 
& \multicolumn{1}{c}{$\varv_{\rm t}^d$} & \multicolumn{1}{c}{$\#^b$} & \multicolumn{1}{c}{WRMS$^e$} \\

\cline{1-6}

0.010~MHz & 3734 & 0.0088~MHz & $\varv_{\rm t}=0 \leftarrow 0$ & 8138 & 0.69 \\
0.020~MHz &  116 & 0.0110~MHz & $\varv_{\rm t}=1 \leftarrow 1$ & 5832 & 0.84 \\
0.030~MHz & 4335 & 0.0223~MHz & $\varv_{\rm t}=2 \leftarrow 2$ &  165 & 0.81 \\
0.040~MHz &    6 & 0.0177~MHz & $\varv_{\rm t}=1 \leftarrow 0$ & 2124 & 0.36 \\
0.050~MHz &  588 & 0.0461~MHz &  &  &  \\
0.100~MHz & 1109 & 0.0779~MHz &  &  &  \\
0.200~MHz &  113 & 0.1662~MHz &  &  &  \\
$5\times 10^{-4}$~cm$^{-1}$ & 4337 & $2.0\times 10^{-4}$~cm$^{-1}$ &   &  & \\ \hline

\end{tabular}
\tablefoot{$^{a}$ Estimated measurement uncertainties for each data group. $^{b}$ Number of lines (left part) 
or transitions (right part) of each category in the least-squares fit. Note that due to blending 14338 
measured line frequencies correspond to 16259  transitions in the fit, which in turn due to presence of 
duplicate measurements represent 15135 unique transitions in the fit.  $^{c}$ Root-mean-square (RMS) deviation 
of corresponding data group. $^{d}$ Upper and lower state torsional quantum number $\varv_{\rm t}$. 
$^{e}$ Weighted root-mean-square (WRMS) deviation of corresponding data group.}
\end{table*}

%%%%%%%%%%%%%%%%%%%%%%%%%%%%%%%%%%%%%%%%%%%%%%%%%%%%%%%%%%%%%%%%%%%%%%%%%%%%%%%%%%%%%%%%%
%%%%%%%%%%%%%%%%%%%%%%%%%%%%%%%%%%%%%%%%%%%%%%%%%%%%%%%%%%%%%%%%%%%%%%%%%%%%%%%%%%%%%%%%%

%%%%%%%%%%%%%%%%%%%%%%%%%%%%%%%%%%%%%%%%%%%%%%%%%%%%%%%%%%%%%%%%%%%%%%%%%%%%%%%%%%%%%%%%%
%%%%%%%%%%%%%%%%%%%%%%%%%%%%%%%%%%%%%%%%%%%%%%%%%%%%%%%%%%%%%%%%%%%%%%%%%%%%%%%%%%%%%%%%%

\onecolumn

\begin{appendix}
\section{Parameters of the RAM Hamiltonian for the CD$_3$OD molecule}

%%%%%%%%%%%%%%%%%%%%%%%%%%%%%%%%%%%%%%%%%%%%%%%%%%%%%%%%%%%%%%%%%%%%%
%%%%%%%%%%%%%%%%%%%%%%%%%%%%%%%%%%%%%%%%%%%%%%%%%%%%%%%%%%%%%%%%%%%%%
%%%%%%    Table 2    %%%%%%%%%%%%%%%%%%%%%%%%%%%%%%%%%%%%%%%%%%%%%%%%
%%%%%%%%%%%%%%%%%%%%%%%%%%%%%%%%%%%%%%%%%%%%%%%%%%%%%%%%%%%%%%%%%%%%%
%%%%%%%%%%%%%%%%%%%%%%%%%%%%%%%%%%%%%%%%%%%%%%%%%%%%%%%%%%%%%%%%%%%%%

\begin{longtable}{lllc}
\caption{\label{tbl:ParametersTable} Fitted parameters of the RAM Hamiltonian for the CD$_3$OD molecule}\\

\hline\hline $n_{tr}$\textit{$^a$} & Operator\textit{$^b$} & Par.\textit{$^{c}$} & Value\textit{$^{d,e}$} \\
\hline
\endfirsthead
\caption{continued.}\\
\hline\hline $n_{tr}$\textit{$^a$} & Operator\textit{$^b$} & Par.\textit{$^{c}$} & Value\textit{$^{d,e}$} \\
\hline
\endhead
\hline

 $ 2_{2,0}$ & $(1-\cos 3\alpha)$                                             & $(1/2)V_3$      &  $  181.1021642(55) $ \\
 $ 2_{2,0}$ & $p_\alpha^2$                                                   & $F$             &  $  14.75939375(90) $ \\
 $ 2_{1,1}$ & $P_ap_\alpha$                                                  & $\rho$          &  $  0.8219648943(19)$ \\
 $ 2_{0,2}$ & $P_a^2$                                                        & $A$             &  $    2.1693625(90) $ \\
 $ 2_{0,2}$ & $P_b^2$                                                        & $B$             &  $    0.63055239(26)$ \\
 $ 2_{0,2}$ & $P_c^2$                                                        & $C$             &  $    0.59843413(22)$ \\
 $ 2_{0,2}$ & $(1/2)\{P_a{,}P_b\}$                                           & $2D_{ab}$       &  $     0.03656819(19)$ \\
 $ 4_{4,0}$ & $(1-\cos 6\alpha)$                                             & $(1/2)V_6$      &  $    -1.076539(13) $ \\
 $ 4_{4,0}$ & $p_\alpha^4$                                                   & $F_m$           &  $    -0.2905843(26) \times 10^{ -2}$ \\
 $ 4_{3,1}$ & $P_ap_\alpha^3$                                                & $\rho_m$        &  $   -0.11365193(86) \times 10^{ -1}$ \\
 $ 4_{2,2}$ & $P^2(1-\cos 3\alpha)$                                          & $V_{3J}$        &  $    -0.1682007(47) \times 10^{ -2}$ \\
 $ 4_{2,2}$ & $P_a^2(1-\cos 3\alpha)$                                        & $V_{3K}$        &  $       0.66900(17) \times 10^{ -2}$ \\
 $ 4_{2,2}$ & $(P_b^2-P_c^2)(1-\cos 3\alpha)$                                & $V_{3bc}$       &  $        0.6739(49) \times 10^{ -5}$ \\
 $ 4_{2,2}$ & $(1/2)\{P_a{,}P_b\}(1-\cos 3\alpha)$                           & $V_{3ab}$       &  $    0.12557689(93) \times 10^{ -1}$ \\
 $ 4_{2,2}$ & $P^2p_\alpha^2$                                                & $F_J$           &  $    -0.5466106(39) \times 10^{ -4}$ \\
 $ 4_{2,2}$ & $P_a^2p_\alpha^2$                                              & $F_K$           &  $    -0.1678373(11) \times 10^{ -1}$ \\
 $ 4_{2,2}$ & $(P_b^2-P_c^2)p_\alpha^2$                                      & $F_{bc}$        &  $     -0.106428(57) \times 10^{ -3}$ \\
 $ 4_{2,2}$ & $(1/2)\{P_a{,}P_c\}\sin 3\alpha$                               & $D_{3ac}$       &  $     0.1808855(27) \times 10^{ -1}$ \\
 $ 4_{2,2}$ & $(1/2)\{P_b{,}P_c\}\sin 3\alpha$                               & $D_{3bc}$       &  $       -0.8227(13) \times 10^{ -3}$ \\
 $ 4_{1,3}$ & $P^2P_ap_\alpha$                                               & $\rho_J$        &  $    -0.8252990(66) \times 10^{ -4}$ \\
 $ 4_{1,3}$ & $P_a^3p_\alpha$                                                & $\rho_K$        &  $   -0.11023649(58) \times 10^{ -1}$ \\
 $ 4_{1,3}$ & $(1/2)\{P_a{,}(P_b^2-P_c^2)\}p_\alpha$                         & $\rho_{bc}$     &  $     -0.151933(58) \times 10^{ -3}$ \\
 $ 4_{1,3}$ & $(1/2)\{P_a^2{,}P_b\}p_\alpha$                                 & $\rho_{ab}$     &  $      -0.22531(72) \times 10^{ -5}$ \\
 $ 4_{0,4}$ & $P^4$                                                          & $-\Delta_J$     &  $    -0.8557802(47) \times 10^{ -6}$ \\
 $ 4_{0,4}$ & $P^2P_a^2$                                                     & $-\Delta_{JK}$  &  $    -0.3430628(28) \times 10^{ -4}$ \\
 $ 4_{0,4}$ & $P_a^4$                                                        & $-\Delta_K$     &  $    -0.2715830(13) \times 10^{ -2}$ \\
 $ 4_{0,4}$ & $P^2(P_b^2-P_c^2)$                                             & $-2\delta_J$    &  $     -0.954346(64) \times 10^{ -7}$ \\
 $ 4_{0,4}$ & $(1/2)\{P_a^2{,}(P_b^2-P_c^2)\}$                               & $-2\delta_K$    &  $     -0.497826(92) \times 10^{ -4}$ \\
 $ 4_{0,4}$ & $(1/2)P^2\{P_a{,}P_b\}$                                        & $D_{abJ}$       &  $      -0.48978(15) \times 10^{ -6}$ \\
 $ 6_{6,0}$ & $p_\alpha^6$                                                   & $F_{mm}$        &  $       0.21266(57) \times 10^{ -5}$ \\
 $ 6_{5,1}$ & $P_ap_\alpha^5$                                                & $\rho_{mm}$     &  $       0.13678(28) \times 10^{ -4}$ \\
 $ 6_{4,2}$ & $P^2(1-\cos 6\alpha)$                                          & $V_{6J}$        &  $        -0.835(18) \times 10^{ -5}$ \\
 $ 6_{4,2}$ & $P_a^2(1-\cos 6\alpha)$                                        & $V_{6K}$        &  $       -0.1930(66) \times 10^{ -3}$ \\
 $ 6_{4,2}$ & $(P_b^2-P_c^2)(1-\cos 6\alpha)$                                & $V_{6bc}$       &  $      -0.31339(65) \times 10^{ -4}$ \\
 $ 6_{4,2}$ & $P^2p_\alpha^4$                                                & $F_{mJ}$        &  $       0.12987(36) \times 10^{ -7}$ \\
 $ 6_{4,2}$ & $P_a^2p_\alpha^4$                                              & $F_{mK}$        &  $       0.36055(57) \times 10^{ -4}$ \\
 $ 6_{4,2}$ & $(1/2)\{P_a{,}P_b\}p_\alpha^4$                                 & $F_{mab}$       &  $        0.3193(70) \times 10^{ -8}$ \\
 $ 6_{4,2}$ & $(1/2)\{P_a{,}P_c\}\sin 6\alpha$                               & $D_{6ac}$       &  $        0.3022(82) \times 10^{ -4}$ \\
 $ 6_{4,2}$ & $(1/2)\{P_b{,}P_c\}\sin 6\alpha$                               & $D_{6bc}$       &  $        0.1206(12) \times 10^{ -4}$ \\
 $ 6_{4,2}$ & $(1/2)\{P_a{,}P_c{,}p_\alpha^2{,}\sin 3\alpha\}$               & $D_{3acm}$      &  $       -0.3795(11) \times 10^{ -4}$ \\
 $ 6_{3,3}$ & $P^2P_ap_\alpha^3$                                             & $\rho_{mJ}$     &  $        0.4567(12) \times 10^{ -7}$ \\
 $ 6_{3,3}$ & $P_a^3p_\alpha^3$                                              & $\rho_{mK}$     &  $       0.50072(62) \times 10^{ -4}$ \\
 $ 6_{3,3}$ & $(1/2)\{P_a^2{,}P_b\}p_\alpha^3$                               & $\rho_{mab}$    &  $        0.2957(67) \times 10^{ -8}$ \\
 $ 6_{3,3}$ & $(1/2)\{P_a{,}P_b{,}P_c{,}p_\alpha{,}\sin 3\alpha\}$           & $\rho_{3bc}$    &  $      -0.26327(58) \times 10^{ -4}$ \\
 $ 6_{2,4}$ & $P^4(1-\cos 3\alpha)$                                          & $V_{3JJ}$       &  $       0.96440(90) \times 10^{ -8}$ \\
 $ 6_{2,4}$ & $P^2P_a^2(1-\cos 3\alpha)$                                     & $V_{3JK}$       &  $      -0.38783(13) \times 10^{ -6}$ \\
 $ 6_{2,4}$ & $P_a^4(1-\cos 3\alpha)$                                        & $V_{3KK}$       &  $       0.43582(61) \times 10^{ -6}$ \\
 $ 6_{2,4}$ & $P^2(P_b^2-P_c^2)(1-\cos 3\alpha)$                             & $V_{3bcJ}$      &  $       0.33135(49) \times 10^{ -8}$ \\
 $ 6_{2,4}$ & $(1/2)\{P_a^2{,}(P_b^2-P_c^2)\}(1-\cos 3\alpha)$               & $V_{3bcK}$      &  $       0.12328(30) \times 10^{ -5}$ \\
 $ 6_{2,4}$ & $(1/2)P^2\{P_a{,}P_b\}(1-\cos 3\alpha)$                        & $V_{3abJ}$      &  $         0.786(13) \times 10^{ -8}$ \\
 $ 6_{2,4}$ & $(1/2)\{P_a^3{,}P_b\}(1-\cos 3\alpha)$                         & $V_{3abK}$      &  $      -0.29046(69) \times 10^{ -5}$ \\
 $ 6_{2,4}$ & $(1/2)(\{P_a{,}P_b^3\}-\{P_a{,}P_b{,}P_c^2\})\cos 3\alpha$     & $V_{3abbc}$     &  $       0.21065(20) \times 10^{ -6}$ \\
 $ 6_{2,4}$ & $P^4p_\alpha^2$                                                & $F_{JJ}$        &  $        0.9587(98) \times 10^{ -9}$ \\
 $ 6_{2,4}$ & $P^2P_a^2p_\alpha^2$                                           & $F_{JK}$        &  $        0.6039(15) \times 10^{ -7}$ \\
 $ 6_{2,4}$ & $P_a^4p_\alpha^2$                                              & $F_{KK}$        &  $       0.38758(38) \times 10^{ -4}$ \\
 $ 6_{2,4}$ & $(1/2)\{P_b^2{,}P_c^2\}p_\alpha^2$                             & $F_{b2c2}$      &  $       -0.4737(77) \times 10^{ -8}$ \\
 $ 6_{2,4}$ & $(1/2)P^2\{P_a{,}P_c\}\sin 3\alpha$                            & $D_{3acJ}$      &  $      -0.32666(10) \times 10^{ -6}$ \\
 $ 6_{2,4}$ & $(1/2)\{P_a^3{,}P_c\}\sin 3\alpha$                             & $D_{3acK}$      &  $       0.24680(74) \times 10^{ -4}$ \\
 $ 6_{2,4}$ & $(1/2)P^2\{P_b{,}P_c\}\sin 3\alpha$                            & $D_{3bcJ}$      &  $       -0.7458(16) \times 10^{ -8}$ \\
 $ 6_{2,4}$ & $(1/2)\{P_a^2{,}P_b{,}P_c\}\sin 3\alpha$                       & $D_{3bcK}$      &  $      -0.20601(44) \times 10^{ -4}$ \\
 $ 6_{2,4}$ & $(1/2)(\{P_a{,}P_c^3\}-\{P_a{,}P_b^2{,}P_c\})\sin 3\alpha$     & $D_{3acbc}$     &  $       0.26239(17) \times 10^{ -6}$ \\
 $ 6_{2,4}$ & $(1/2)(\{P_b^3{,}P_c\}-\{P_b{,}P_c^3\})\sin 3\alpha$           & $D_{3bcbc}$     &  $        -0.345(22) \times 10^{ -8}$ \\
 $ 6_{1,5}$ & $P^4P_ap_\alpha$                                               & $\rho_{JJ}$     &  $        0.9957(92) \times 10^{ -9}$ \\
 $ 6_{1,5}$ & $P^2P_a^3p_\alpha$                                             & $\rho_{JK}$     &  $       0.36124(82) \times 10^{ -7}$ \\
 $ 6_{1,5}$ & $P_a^5p_\alpha$                                                & $\rho_{KK}$     &  $       0.15889(13) \times 10^{ -4}$ \\
 $ 6_{1,5}$ & $(1/2)P^2\{P_a{,}(P_b^2-P_c^2)\}p_\alpha$                      & $\rho_{bcJ}$    &  $        0.3819(75) \times 10^{ -9}$ \\
 $ 6_{1,5}$ & $(1/2)\{P_a{,}P_b^2{,}P_c^2\}p_\alpha$                         & $\rho_{b2c2}$   &  $       -0.4311(73) \times 10^{ -8}$ \\
 $ 6_{0,6}$ & $P^6$                                                          & $\Phi_J$        &  $        0.4568(18) \times 10^{-12}$ \\
 $ 6_{0,6}$ & $P^4P_a^2$                                                     & $\Phi_{JK}$     &  $       0.13849(13) \times 10^{ -9}$ \\
 $ 6_{0,6}$ & $P^2P_a^4$                                                     & $\Phi_{KJ}$     &  $        0.8455(18) \times 10^{ -8}$ \\
 $ 6_{0,6}$ & $P_a^6$                                                        & $\Phi_K$        &  $       0.26994(18) \times 10^{ -5}$ \\
 $ 6_{0,6}$ & $P^4(P_b^2-P_c^2)$                                             & $2\phi_J$       &  $       0.43864(71) \times 10^{-12}$ \\
 $ 6_{0,6}$ & $(1/2)P^2\{P_a^2{,}(P_b^2-P_c^2)\}$                            & $2\phi_{JK}$    &  $        0.3821(72) \times 10^{ -9}$ \\
 $ 6_{0,6}$ & $(1/2)P^4\{P_a{,}P_b\}$                                        & $D_{abJJ}$      &  $         0.901(12) \times 10^{-12}$ \\
 $ 8_{8,0}$ & $p_\alpha^8$                                                   & $F_{mmm}$       &  $        0.4599(43) \times 10^{ -9}$ \\
 $ 8_{7,1}$ & $P_ap_\alpha^7$                                                & $\rho_{mmm}$    &  $        0.1264(12) \times 10^{ -8}$ \\
 $ 8_{6,2}$ & $P^2(1-\cos 9\alpha)$                                          & $V_{9J}$        &  $        0.1612(37) \times 10^{ -4}$ \\
 $ 8_{6,2}$ & $P_a^2(1-\cos 9\alpha)$                                        & $V_{9K}$        &  $         0.293(14) \times 10^{ -3}$ \\
 $ 8_{6,2}$ & $(P_b^2-P_c^2)(1-\cos 9\alpha)$                                & $V_{9bc}$       &  $        0.4795(74) \times 10^{ -5}$ \\
 $ 8_{6,2}$ & $P_a^2p_\alpha^6$                                              & $F_{mmK}$       &  $        0.9527(87) \times 10^{ -9}$ \\
 $ 8_{6,2}$ & $(1/2)\{P_a{,}P_c\}\sin 9\alpha$                               & $D_{9ac}$       &  $        -0.764(24) \times 10^{ -4}$ \\
 $ 8_{6,2}$ & $(1/2)\{P_a{,}P_c{,}p_\alpha^2{,}\sin 6\alpha\}$               & $D_{6acm}$      &  $         0.478(13) \times 10^{ -6}$ \\
 $ 8_{4,4}$ & $P^4(1-\cos 6\alpha)$                                          & $V_{6JJ}$       &  $        0.6631(69) \times 10^{ -9}$ \\
 $ 8_{4,4}$ & $P_a^4(1-\cos 6\alpha)$                                        & $V_{6KK}$       &  $        -0.382(10) \times 10^{ -7}$ \\
 $ 8_{4,4}$ & $P^2(P_b^2-P_c^2)(1-\cos 6\alpha)$                             & $V_{6bcJ}$      &  $       0.11622(71) \times 10^{ -8}$ \\
 $ 8_{4,4}$ & $(1/2)\{P_a^3{,}P_b\}(1-\cos 6\alpha)$                         & $V_{6abK}$      &  $        -0.432(11) \times 10^{ -7}$ \\
 $ 8_{3,5}$ & $(1/2)\{P_a^3{,}P_b{,}P_c{,}p_\alpha{,}\sin 3\alpha\}$         & $\rho_{3bcK}$   &  $        0.1901(18) \times 10^{ -8}$ \\
 $ 8_{2,6}$ & $P^6(1-\cos 3\alpha)$                                          & $V_{3JJJ}$      &  $       -0.1652(40) \times 10^{-12}$ \\
 $ 8_{2,6}$ & $P^4P_a^2(1-\cos 3\alpha)$                                     & $V_{3JJK}$      &  $        0.7750(28) \times 10^{-11}$ \\
 $ 8_{2,6}$ & $P_a^6(1-\cos 3\alpha)$                                        & $V_{3KKK}$      &  $         0.829(47) \times 10^{-10}$ \\
 $ 8_{2,6}$ & $P^4(P_b^2-P_c^2)(1-\cos 3\alpha)$                             & $V_{3bcJJ}$     &  $       -0.2301(17) \times 10^{-12}$ \\
 $ 8_{2,6}$ & $(1/2)P^2\{P_a^2{,}(P_b^2-P_c^2)\}(1-\cos 3\alpha)$            & $V_{3bcJK}$     &  $        0.1481(89) \times 10^{-11}$ \\
 $ 8_{2,6}$ & $(1/2)P^2\{P_b^2{,}P_c^2\}\cos 3\alpha$                        & $V_{3b2c2J}$    &  $        -0.756(32) \times 10^{-12}$ \\
 $ 8_{2,6}$ & $(1/2)(\{P_b^4{,}P_c^2\}-\{P_b^2{,}P_c^4\})\cos 3\alpha$       & $V_{3b2c2bc}$   &  $      -0.22643(76) \times 10^{-11}$ \\
 $ 8_{2,6}$ & $(1/2)P^2(\{P_a{,}P_b^3\}-\{P_a{,}P_b{,}P_c^2\})\cos 3\alpha$  & $V_{3abbcJ}$    &  $        -0.682(47) \times 10^{-12}$ \\
 $ 8_{2,6}$ & $P^6p_\alpha^2$                                                & $F_{JJJ}$       &  $        0.5680(96) \times 10^{-15}$ \\
 $ 8_{2,6}$ & $P_a^6p_\alpha^2$                                              & $F_{KKK}$       &  $       -0.5406(46) \times 10^{ -9}$ \\
 $ 8_{2,6}$ & $P^4(P_b^2-P_c^2)p_\alpha^2$                                   & $F_{bcJJ}$      &  $        -0.950(50) \times 10^{-15}$ \\
 $ 8_{2,6}$ & $(1/2)(\{P_b^4{,}P_c^2\}-\{P_b^2{,}P_c^4\})p_\alpha^2$         & $F_{b2c2bc}$    &  $        0.1125(40) \times 10^{-13}$ \\
 $ 8_{2,6}$ & $(1/2)P^4\{P_a{,}P_c\}\sin 3\alpha$                            & $D_{3acJJ}$     &  $        0.4398(52) \times 10^{-11}$ \\
 $ 8_{2,6}$ & $(1/2)P^2\{P_a^3{,}P_c\}\sin 3\alpha$                          & $D_{3acJK}$     &  $        0.1247(30) \times 10^{-10}$ \\
 $ 8_{2,6}$ & $(1/2)P^4\{P_b{,}P_c\}\sin 3\alpha$                            & $D_{3bcJJ}$     &  $       0.10470(39) \times 10^{-11}$ \\
 $ 8_{2,6}$ & $(1/2)\{P_a^4{,}P_b{,}P_c\}\sin 3\alpha$                       & $D_{3bcKK}$     &  $        0.1406(14) \times 10^{ -8}$ \\
 $ 8_{2,6}$ & $(1/2)P^2(\{P_a{,}P_c^3\}-\{P_a{,}P_b^2{,}P_c\})\sin 3\alpha$  & $D_{3acbcJ}$    &  $       -0.5366(44) \times 10^{-11}$ \\
 $ 8_{2,6}$ & $(1/2)P^2(\{P_b^3{,}P_c\}-\{P_b{,}P_c^3\})\sin 3\alpha$        & $D_{3bcbcJ}$    &  $        0.3186(95) \times 10^{-12}$ \\
 $ 8_{2,6}$ & $(1/2)\{P_b^3{,}P_c^3\}\sin 3\alpha$                           & $D_{3b3c3}$     &  $       -0.4538(18) \times 10^{-11}$ \\
 $ 8_{1,7}$ & $P_a^7p_\alpha$                                                & $\rho_{KKK}$    &  $       -0.5405(45) \times 10^{ -9}$ \\
 $ 8_{0,8}$ & $P_a^8$                                                        & $L_K$           &  $       -0.1484(12) \times 10^{ -9}$ \\
 $10_{8,2}$ & $(1/2)\{P_a{,}P_c{,}p_\alpha^2{,}\sin 9\alpha\}$               & $D_{9acm}$      &  $        -0.662(23) \times 10^{ -6}$ \\
 $10_{6,4}$ & $(1/2)\{P_a^2{,}(P_b^2-P_c^2)\}(1-\cos 9\alpha)$               & $V_{9bcK}$      &  $       -0.4026(56) \times 10^{ -7}$ \\
 $10_{4,6}$ & $P_a^6(1-\cos 6\alpha)$                                        & $V_{6KKK}$      &  $        -0.885(51) \times 10^{-10}$ \\
 $10_{4,6}$ & $(1/2)P^4\{P_a{,}P_c\}\sin 6\alpha$                            & $D_{6acJJ}$     &  $       -0.1390(70) \times 10^{-11}$ \\
 $10_{4,6}$ & $(1/2)P^2\{P_a^2{,}P_b{,}P_c\}\sin 6\alpha$                    & $D_{6bcJK}$     &  $       -0.1579(29) \times 10^{-10}$ \\
 $10_{4,6}$ & $(1/2)P^2(\{P_a{,}P_c^3\}-\{P_a{,}P_b^2{,}P_c\})\sin 6\alpha$  & $D_{6acbcJ}$    &  $         0.595(13) \times 10^{-11}$ \\
 $10_{2,8}$ & $P_a^8(1-\cos 3\alpha)$                                        & $V_{3KKKK}$     &  $        -0.217(15) \times 10^{-12}$ \\
 $10_{2,8}$ & $(1/2)(\{P_b^6{,}P_c^2\}+\{P_b^2{,}P_c^6\})\cos 3\alpha$       & $V_{3b6c2b2c6}$ &  $         0.304(18) \times 10^{-16}$ \\
 $12_{4,8}$ & $P_a^8(1-\cos 6\alpha)$                                        & $V_{6KKKK}$     &  $         0.222(17) \times 10^{-12}$ \\

\hline
\hline

\end{longtable}

\tablefoot{$^{a}$ \textit{n=t+r}, where \textit{n} is the total order of the operator, \textit{t} is the order 
  of the torsional part and \textit{r} is the order of the rotational part, respectively. The ordering scheme 
  of \citet{Nakagawa:1987} is used. $^{b}$ $\lbrace A,B,C,D,E \rbrace = ABCDE+EDCBA$.  $\lbrace A,B,C,D \rbrace = ABCD+DCBA$. 
  $\lbrace A,B,C \rbrace = ABC+CBA$. $\lbrace A,B \rbrace = AB+BA$. 
  The product of the operator in the second column of a given row and the parameter in the third column 
  of that row gives the term actually used in the torsion-rotation Hamiltonian of the program, except for 
  \textit{F}, $\rho$ and \textit{$A_{\rm RAM}$}, which occur in the Hamiltonian in the form 
  $F(p_a + \rho P_a)^2 + A_{\rm RAM}P_a^2$. $^{c}$ The parameter nomenclature is based on the subscript 
  procedure of \citet{XU:2008305}. $^{d}$ Values of the parameters in units of cm$^{-1}$, except for $\rho$, 
  which is unitless. $^{e}$ Statistical uncertainties are given in parentheses as one standard uncertainty 
  in units of the last digits.}

%%%%%%%%%%%%%%%%%%%%%%%%%%%%%%%%%%%%%%%%%%%%%%%%%%%%%%%%%%%%%%%%%%%%%%%%%%%%%%%%%%%%%%%%%%
%%%%%%%%%%%%%%%%%%%%%%%%%%%%%%%%%%%%%%%%%%%%%%%%%%%%%%%%%%%%%%%%%%%%%%%%%%%%%%%%%%%%%%%%%%

\end{appendix}

\end{document}